\documentstyle[multicol,aps,epsf]{revtex}
\begin{document}
\draft
\title{Transverse force on a quantized vortex in a superconductor} 
\author{Michael R. Geller,$^{1,2}$ Carlos Wexler,$^{1,3}$ and David J. Thouless$^1$}
\address{$^1$Department of Physics, University of Washington, Seattle, Washington 98195}
\address{$^2$Department of Physics and Astronomy, University of Georgia, Athens, Georgia 30602}
\address{$^3$Department of Physics, University of Florida, Gainesville, FL 32611}
\date{\today}
\maketitle

\begin{abstract}
The total transverse force acting on a quantized vortex in a type-II 
superconductor determines the Hall response in the mixed state, 
yet a consensus as to its correct form is still lacking. In this 
paper we present an essentially exact expression for this force,
valid in the superclean limit, which was obtained by generalizing 
the recent work by Thouless, Ao, and Niu [D. J. Thouless, 
P. Ao, and Q. Niu, Phys.~Rev.~Lett. {\bf 76}, 3758 (1996)] on the 
Magnus force in a neutral superfluid. We find the transverse force per 
unit length to be ${\bf f} = \rho \; {\bf K} \times {\bf V}$, where 
$\rho = \rho_{\rm n} + \rho_{\rm s}$ is the sum of the mass densities
of the normal and superconducting components, 
${\bf K}$ is a vector parallel to the line vortex
with a magnitude equal to the quantized circulation, and ${\bf V}$ 
is the vortex velocity. 
\end{abstract}

\begin{multicols}{2}

The discovery of high-temperature superconductivity has stimulated
a renewed interest in the statistical mechanics and dynamics of
vortices in type-II superconductors \cite{Blatter review}. Despite the 
tremendous progress made, however, the answer to one of the simplest 
questions, namely, the form of the equation of motion for a single
isolated vortex, remains controversial. If we let ${\bf R}(t)$ denote
the position in the $xy$ plane of the center of a straight vortex line 
as a function of time, then the classical equation of motion can be 
written as  
\begin{equation}
\label{equation of motion}
M \: {d^2 {\bf R} \over dt^2} = - \eta \: {d {\bf R} \over dt} 
- \gamma \: {d {\bf R} \over dt} \times {\bf e}_z 
+ {\bf f}_{\rm p}({\bf R}) + {\bf f}_{\rm d}({\bf R}).
\end{equation}
Here we have taken the circulation vector ${\bf K}$ of the vortex (a vector 
parallel to the vortex with a magnitude equal to the 
circulation \cite{circulation footnote}) to be along the $z$ direction. 
The first two terms on the right-hand-side of (\ref{equation of motion}) are 
to include all forces linear in the vortex velocity. Here ${\bf f}_{\rm p}$ 
represents the pinning force due to disorder (averaged over an area set by 
the size of the vortex core), and ${\bf f}_{\rm d}$ includes the various ``driving'' 
forces possible, such as the Lorentz force, which may depend on the density of the 
normal and superconducting components $n_{\rm s}$ and $n_{\rm n}$, and 
therefore on the vortex position ${\bf R}$, but do not depend on the vortex 
velocity. The equation of motion (\ref{equation of motion}) also describes a
vortex in a Bose or Fermi superfluid, where ${\bf f}_{\rm p}$ might describe
the force from an externally imposed wire as in a Vinen-like experiment \cite{Vinen}, 
and ${\bf f}_{\rm d}$ would include the superfluid-velocity-dependent
part of the Magnus force \cite{Wexler} and possibly other vortex-velocity-independent 
contributions \cite{Sonin}.

The difficulty concerns the determination of the coefficients 
$M$, $\eta$, and $\gamma$,
which describe the vortex effective mass per unit length, viscous 
damping force per unit length, and nondissipative transverse force 
per unit length, respectively. 
Of particular interest and especially controversial is the  
coefficient $\gamma$, which in a neutral superfluid has
been recently shown by Thouless, Ao, and Niu \cite{TAN} (hereafter 
referred to as TAN) to be $ \gamma = \rho_{\rm s} \:\! K $, 
where $\rho_{\rm s}$ is the mass density of the superfluid component 
far from the vortex core, and $K$ is the quantized circulation. 
For a vortex in a neutral Bose superfluid with $q$ quanta of circulation, 
this coefficient is simply $\gamma = q h n_{\rm s}$, 
with $n_{\rm s}$ the superfluid number density away from the core, 
whereas in a (paired) Fermi superfluid $\gamma = {1 \over 2}\:\! q \:\! h \:\! n_{\rm s}$.
The TAN result shows that the transverse 
force is {\it universal}, independent of the detailed microscopic 
structure of the vortex and its interaction with the normal component
of the fluid.
The universality of the nondissipative transverse force is not found 
in some approximate microscopic calculations, however, which obtain 
coefficients that reduce to the TAN result only in certain limits
\cite{Kopnin,Volovik,van Otterlo etal}.    

In this paper we calculate the nondissipative transverse force on a quantized
vortex in a charged superfluid or superconductor by following the method of
TAN \cite{TAN}. Initially, we shall consider a system with
translational invariance, and to address a more realistic model of a 
superconductor we include afterwards a periodic potential from the 
lattice, assuming the lattice constant $b$ is small compared with the 
coherence length $\xi$. In both cases we find the transverse force
per length to be given by
\begin{equation}
\label{transverse force}
{\bf f} = \rho \; {\bf K} \times {\bf V},
\end{equation}
where $\rho = \rho_{\rm n} + \rho_{\rm s}$ is the {\it total} mass
density of the fluid away from the core, ${\bf K}$ is the quantized 
circulation vector defined above, and ${\bf V}$ is the constant vortex 
velocity. 
Most relevant is the fact that, as in TAN, our results for this 
superclean limit of a superconductor also indicate a 
{\it universal} vortex-velocity-dependent part of the transverse force.
In the notation of Eqn.~(\ref{equation of motion}), we find
$\gamma = \rho \:\! K$. For a vortex with $q$ trapped flux quanta
$\Phi_0 \equiv hc/2e$ we can write this as 
$\gamma = {1 \over 2} \:\! q \:\! h \:\! n$,
where $n$ is the total density of the fluid far from the core. 
The reason that the transverse force in the charged case differs 
from that for the neutral superfluid, which has a coefficient 
$\rho_{\rm s}$ instead of $\rho$, is related to the Meissner effect, 
as will be explained below.

The Hamiltonian for the charged superfluid or superconductor is taken to be
\begin{equation}
\label{full Hamiltonian} 
H = H_0 + H_1 + \sum_n V({\bf r}_n - {\bf R}),
\end{equation}
where 
\begin{equation}
\label{fluid hamiltonian}
H_0 \equiv \sum_n {p_n^2 \over 2m} 
+ {1 \over 2} \sum_{n \neq n'} U({\bf r}_n 
- {\bf r}_{n'}) + H_{\rm b}
\end{equation}
is the Hamiltonian for $N$ bosons or fermions with negative charge $-e$,
interacting with each other and with a uniform positive background charge, and
(in Gaussian units)
\begin{equation}
\label{current-current interaction}
H_1 \equiv - {e^2 \over 2 m^2 c} \sum_{n \neq n'}
p_n^i \ \! T^{ij}({\bf r}_n - {\bf r}_{n'}) \ \! p_{n'}^j,
\end{equation} 
with $T^{ij}({\bf r}) \equiv (\delta^{ij} |{\bf r}|^{-1} 
+ r^ir^j |{\bf r}|^{-3})/2c,$ is the
current-current interaction, which, as first noted by Darwin 
\cite{Jackson}, correctly accounts for the electrodynamics
in the transverse gauge through order $v^2/c^2$.
The $U$ in (\ref{fluid hamiltonian}) is a Coulomb interaction term, and
in the case of a superconductor also contains an additional 
short-ranged attractive interaction to produce superconductivity. 
Also, $H_{\rm b} \equiv - \sum_n \int d^3r \ \! e^2 {\bar n}
|{\bf r}_n -{\bf r}|^{-1}$ accounts for the interaction with the
uniform positive background charge $e {\bar n}$. 
The full Hamiltonian also contains a 
cylindrically symmetric pinning potential $V$ centered at 
position ${\bf R}$ \cite{relativistic footnote}. 

Before proceeding it is important to note that the Hamiltonian
(\ref{full Hamiltonian}), with the current-current interaction term, provides
an accurate model of a charged superfluid or superconductor. In particular, 
we show in the appendix that
it exhibits a Meissner effect with the conventional London screening 
length.

We shall follow TAN and calculate the force on the vortex by expanding the 
time-dependent wave function, given by $ i\hbar \partial_t | 
\Psi(t) \rangle = H\big({\bf R}(t) \big) | \Psi(t) \rangle,$ 
in a basis of instantaneous eigenstates satisfying 
$ H({\bf R}) \big|\psi_\alpha ({\bf R}) 
\big\rangle = E_\alpha ({\bf R}) \big|\psi_\alpha ({\bf R}) \big\rangle . $ 
We shall assume a three-dimensional system with the $z$ direction along 
${\bf K}$. The system is assumed to be infinite in the $x$ and $y$ 
directions, and $L$ is the thickness in the $z$ direction, which is 
also the length of the vortex. In the absence of the pinning potential
the system is therefore translationally invariant in the $x$ and $y$ 
directions.  The translational invariance allows the instantaneous 
eigenfunctions to be taken as $ \psi_\alpha^{\bf R}({\bf r}_1, 
\cdots , {\bf r}_N ) = \psi_\alpha({\bf r}_1 - {\bf R}, \cdots , 
{\bf r}_N - {\bf R} ),$ where $\psi_\alpha({\bf r}_1, \cdots ,{\bf r}_N )$ 
are the eigenfunctions with the pinning potential centered at the origin. 
Initially, at a time $t_0$, a vortex is assumed to be bound to 
the pinning potential at ${\bf R}(t_0) = {\bf R}_0$ in a state characterized 
by the density matrix $\sum_\alpha f_\alpha |\psi_\alpha 
({\bf R}_0) \rangle \langle \psi_\alpha ({\bf R}_0) |,$ where 
$f_\alpha$ is the  occupation probability of state $\alpha$. The force 
on the vortex is $ {\bf F} = - \sum_\alpha f_\alpha \ \langle 
\Psi_\alpha (t) |\nabla_R H |\Psi_\alpha (t) \rangle,$
where $|\Psi_\alpha (t) \rangle$ is the solution of the time-dependent 
Schr\"odinger equation starting out in 
$|\psi_\alpha ({\bf R}(t_0) ) \rangle$. Following Ref.~\onlinecite{TAN} 
we obtain a transverse force equal to
\begin{equation}
{\bf F} = {i \hbar L \over 2} \: ({\bf e}_z \times {\bf V}) \:
\oint_a d{\bf l} \cdot \big(\nabla - \nabla' \big) 
\rho({\bf r},{\bf r}') \big|_{{\bf r}' \rightarrow {\bf r}},
\label{general transverse force}
\end{equation}
where $\rho({\bf r},{\bf r}')$ is the one-particle density matrix and 
$L$ is the thickness of the system in the $z$ direction. In 
(\ref{general transverse force}) we choose the radius $a$ of the 
integration contour to be much 
larger than the London penetration depth $\lambda$.

The integrand in (\ref{general transverse force}) is proportional
to the canonical momentum density. The current-current interaction
term (\ref{current-current interaction}) has played no role
up to this point, and, indeed, the expression (\ref{general
transverse force}) in the two-dimensional limit is identical to that 
obtained by TAN for the neutral superfluid. However, the resulting 
force is different, as can be seen by writing 
(\ref{general transverse force}) in terms of the canonical 
momentum density, which leads to a transverse force per unit length equal to
\begin{equation}
{\bf f} = {\bf e}_z \times {\bf V} \: 
\oint_a d{\bf l}   \cdot {\bf j}({\bf r}).
\label{simplified force}
\end{equation}
The TAN result for neutral superfluids is obtained by writing 
the momentum density as 
${\bf j} = \rho_{\rm n} {\bf v}_{\rm n} + \rho_{\rm s} {\bf v}_{\rm s}$
and assuming that the viscous normal component does not 
circulate \cite{Wexler and Thouless}.
In the case of a charged superfluid or superconductor, however,
the gauge-invariant momentum density ${\bf j} + {e \over c} n {\bf A}$
vanishes in the region containing the integration contour because 
of the Meissner effect (see Appendix), and we therefore 
obtain (\ref{transverse force}). 
 
Our conclusion suggests that both the normal
and superconducting  components of the fluid contribute to the total
transverse force. However, we would like  to emphasize that the model
considered here does not include disorder, and, as such, leads  to a
normal component having an infinite conductivity. This neglect of
disorder may be  responsible for the $\rho$ dependence of the
transverse force instead of the usual  $\rho_{\rm s}$.

Finally, we would like to remark that the result 
(\ref{transverse force}) holds even in the presence of a lattice,
as long as the coherence length or vortex core size $\xi$
is large compared with the lattice constant $b$. To demonstrate
this we add a periodic potential term
$\sum_n v({\bf r}_n)$ to $H_0$ and expand the instantaneous
eigenstates in a basis of localized Wannier functions
$a({\bf r-l})$ for the relevant band,
\begin{eqnarray}
  \psi_\alpha^{\bf R}({\bf r}_1, \cdots, {\bf r}_N)
  & = & \sum_{{\bf l}_1 \cdots {\bf l}_N} 
  \phi_\alpha^{\bf R}({\bf l}_1, \cdots, {\bf l}_N)
  \ \! a({\bf r}_1 - {\bf l}_1) \nonumber \\
  &\times& a({\bf r}_2 - {\bf l}_2) \cdots a({\bf r}_N - {\bf l}_N).
\end{eqnarray}
Here ${\bf l}$ labels the sites of the lattice, and the
coefficients $\phi_\alpha^{\bf R}({\bf l}_1, \cdots, {\bf l}_N)$
are taken to be completely symmetric or antisymmetric.
Assuming that we are at the minimum of a band with an 
isotropic effective mass $m^*$, we see that the envelope function
$\phi_\alpha^{\bf R}({\bf r}_1, \cdots, {\bf r}_N)$ satisfies
a Schr\"odinger equation with a Hamiltonian given by
(\ref{full Hamiltonian}) apart from the replacement
of $m$ in $H_0$ with $m^*$ (the mass in $H_1$ is
not changed). This is the standard effective mass 
approximation, and it leads to
\begin{equation}
  {\bf F} = {i \hbar \over 2}  ({\bf e}_z \times {\bf V}) 
  \int_0^L dz \oint_a d{\bf l} \cdot \big(\nabla - \nabla' \big) 
  \rho_{\rm eff}({\bf r},{\bf r}') \big|_{{\bf r}' \rightarrow {\bf r}},
  \label{effective transverse force}
\end{equation}
where
\begin{eqnarray}
  \rho_{\rm eff}({\bf r},{\bf r}') & \equiv & N \sum_\alpha 
  f_\alpha  \int  d^3 r_2 
  d^3r_3 \cdots d^3r_N \nonumber \\
  & \times & \phi_\alpha^*({\bf r}, {\bf r}_2, \cdots, {\bf r}_N ) 
  \ \! \phi_\alpha ({\bf r}', {\bf r}_2, \cdots , {\bf r}_N ) 
\end{eqnarray} 
is a one-particle density matrix constructed from the envelope 
functions.

Unfortunately, the integrand in (\ref{effective transverse force})
is {\it not} proportional to the actual canonical momentum density 
${\bf j}({\bf r})$, the latter having fluctuations on the 
scale of the lattice constant $b$. However, it is possible 
to prove that when any {\it local} single-particle operator 
like the current density is averaged over a length scale 
larger than the localization length of the Wannier functions 
(assumed to be of the order of $b$), but smaller than the 
characteristic length scale over which the envelope 
functions vary, the expectation value is correctly
given by $\rho_{\rm eff}({\bf r},{\bf r}')$ 
\cite{coarse-graining footnote}. In other words, the structure on 
the scale of the lattice constant is not described correctly
by the envelope functions, but coarse-grained quantities are.   
Now, because the line integral of the actual canonical momentum 
is quantized, it is possible to write it as
\begin{equation}
  \oint_a d{\bf l} \cdot {\bf j} = {1 \over a_2 - a_1} \int_{a_1}^{a_2} da 
  \ {1 \over L} \int_0^L dz \oint_a d{\bf l} \cdot {\bf j},
  \label{averaged circulation}
\end{equation}
where both radii $a_1$ and $a_2$ are much larger than $\lambda$
and their difference is larger than $b$. The integrals over
$a$ and $z$ on the right-hand-side of (\ref{averaged circulation}) 
have the effect of averaging the azimuthal component of ${\bf j}$. 
Therefore, the actual circulation
is correctly given by the line integral of the coarse-grained
momentum, which, in turn, is correctly given by the
envelope functions. Hence, (\ref{effective transverse force})
leads to the transverse force (\ref{transverse force}), as stated.

The total transverse force per unit length on the vortex may be obtained by adding the conventional
Lorentz force term ${\bf f}_{\rm L} = - {e \over c} \ \!  n \ \! {\bf v}_{\rm e} \times {\bf \Phi},$ leading to 
\end{multicols}
\begin{equation}
{\bf f} = \rho {\bf K} \times ( {\bf V} - {\bf v}_{\rm e} )
= {e \over c} \int d^2r \ n  \ \!  ( {\bf V} - {\bf v}_{\rm e} ) \times {\bf B}.
\label{galilean invariant force}
\end{equation}
Alternatively, we can write this as
\begin{equation}
{\bf f} = {e \over c} \int d^2r \  n \ \! ( {\bf v}_{\rm p} - {\bf v}_{\rm e} ) \times {\bf B} 
+ {e \over c} \int d^2r \  n  \ \! ( {\bf V} -  {\bf v}_{\rm p} ) \times {\bf B}  ,
\label{alternate form}
\end{equation}
where ${\bf v}_{\rm p}$ is the velocity of the positively charged substrate, usually taken to be at rest.
The first term in (\ref{alternate form}) is the Lorentz force, given by the interaction of the 
{\it Galilean-invariant} current with the magnetic field, while the second term is a Magnus force 
that acts on the substrate. This interpretation is in agreement with the early picture of flux line motion
in type-II superconductors given by Nozi\`eres and Vinen in the late sixties \cite{Nozieres and Vinen}: 
The Magnus force on the vortex may be thought as the Kutta-Joukowski hydrodynamic lift force due to the
circulation of the electron fluid around the vortex. Far from the vortex the circulation is reduced but an 
increase in the Lorentz force exactly compensates this deficiency. The Magnus force reaction is
eventually carried away by the positive substrate.

\begin{multicols}{2}

This work was supported by the NSERC of Canada and by the NSF through grant No. DMR-9528345. 
M.~G. would like to acknowledge the kind hospitality of the Department of Physics at the University
of Washington where this work was done.

\appendix
\section{Meissner effect in the RPA approximation}

As mentioned above, it is important to establish that the Darwin Hamiltonian provides
an accurate model of a charged superfluid or superconductor;
in particular, that it exhibits a Meissner effect with the 
conventional London penetration depth. 

\begin{figure}
\begin{center}       
\leavevmode
\epsfbox{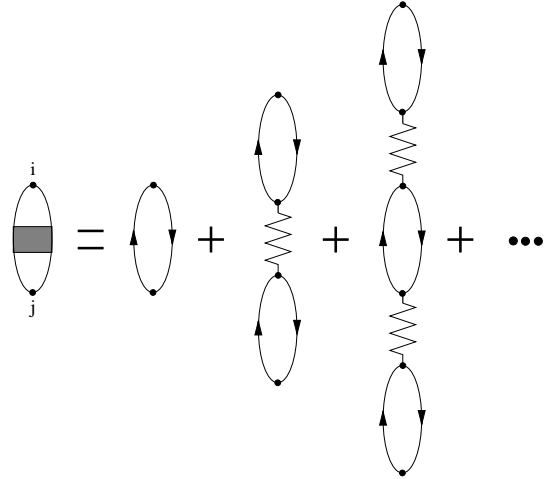}
\end{center}
\caption{RPA approximation for the correlation function $\Pi^{ij}(i \omega_n, {\bf q})$. Here the solid lines 
denote the exact Green's functions for $H_0$ and the zig-zag lines represent the current-current interaction 
$T^{ij}({\bf q})$.}
\label{RPA}
\end{figure}

This can be demonstrated by treating the current-current interactions 
in an RPA-type approximation and calculating the current induced by 
a weak applied vector potential,
\begin{equation}
J^i({\bf q}) = \chi^{ij}({\bf q}) A_{\rm ext}^j({\bf q}).
\end{equation}
The zero-frequency linear response function $\chi^{ij}$
is the sum of a retarded current-current correlation
function $\Pi_{\rm R}^{ij}$ for a system described by the 
Hamiltonian $H_0 + H_1$, and a diamagnetic term.
Our RPA approximation corresponds to a summation of the 
diagrams shown in Fig.~\ref{RPA} for the imaginary-time
current-current correlation function $\Pi^{ij}(i \omega_n,{\bf q})$,
which are the most divergent terms as ${\bf q} \rightarrow 0$. 
In this approximation we find
\begin{equation}
\chi = \chi_0 + \chi_0 \ \! T \ \! \chi,
\label{response function}
\end{equation}
where $\chi_0^{ij}$ is the corresponding response function
for the system without current-current interactions,
as described by $H_0$, and
$T^{ij}({\bf q}) = 4 \pi (\delta^{ij} |{\bf q}|^{-2} 
- q^i q^j |{\bf q}|^{-4})/c$ is the Fourier transform
of $T^{ij}({\bf r})$.

The response function $\chi^{ij}$ can be used to
relate the total vector potential ${\bf A}_{\rm tot} \equiv
{\bf A}_{\rm ext} + {\bf A}_{\rm ind}$, the sum of an external
and induced part, to  ${\bf A}_{\rm ext}$ itself,
\begin{equation}
A_{\rm tot}^i({\bf q}) = [1 - T({\bf q}) \chi_0({\bf q})]^{-1}_{ij}
A_{\rm ext}^j({\bf q}).
\label{magnetic screening}
\end{equation}
Because $\omega = 0$ here, (\ref{magnetic screening}) tells us 
about {\it magnetic} screening. Now, $\chi_0^{ij}$
is known to have some general properties in the 
${\bf q} \rightarrow 0$ limit, reflecting the presence of
off-diagonal long-range order. In particular,
\begin{equation}
\chi_0^{ij}({\bf q}) = {n_{\rm s} e^2 \over mc}  
\big( \delta^{ij} - q^i q^j / |{\bf q}|^{2} \big)
\label{ODLRO}
\end{equation}
in this limit \cite{Schrieffer}, resulting in
\begin{equation}
{\bf A}_{\rm tot}({\bf q}) = \bigg(1 + {1 \over \lambda^2
q^2} \bigg)^{-1} {\bf A}_{\rm ext}({\bf q}),
\label{Meissner effect} 
\end{equation}
where $\lambda \equiv (m c^2 / 4 \pi n_{\rm s} e^2 )^{1 \over 2}$
is the usual London penetration depth. Thus, the current-current
interaction term leads to a conventional Meissner effect.

\end{multicols}

\begin{references}
\bibitem{relativistic footnote} The other order $v^2/c^2$ relativistic corrections do not affect our result 
(\ref{transverse force}).
\end{references}
\end{document}